\documentclass[12pt,a4paper,superscriptaddress]{revtex4}
\usepackage{graphicx}
\usepackage{graphics}
\usepackage{amsmath}
\usepackage{dcolumn}
\usepackage{amssymb}
\usepackage{bm}
\usepackage[latin1]{inputenc}
\newcommand{\be}{\begin{equation}}
\newcommand{\ee}{\end{equation}}
\newcommand{\bea}{\begin{eqnarray}}
\newcommand{\eea}{\end{eqnarray}}

\newcommand{\pa}{\partial}

\begin{document}

\title{On the effective potential for Horava-Lifshitz-like theories with the arbitrary critical exponent}

\author{C. F. Farias}

\affiliation{Departamento de F\'{\i}sica, Universidade Federal da 
Para\'{\i}ba\\
Caixa Postal 5008, 58051-970, Jo\~ao Pessoa, Para\'{\i}ba, Brazil}
\email{cffarias,jroberto,petrov@fisica.ufpb.br}

\author{J. R. Nascimento}

\affiliation{Departamento de F\'{\i}sica, Universidade Federal da
Para\'{\i}ba\\
Caixa Postal 5008, 58051-970, Jo\~ao Pessoa, Para\'{\i}ba, Brazil}
\email{cffarias,jroberto,petrov@fisica.ufpb.br}

\author{A. Yu. Petrov}

\affiliation{Departamento de F\'{\i}sica, Universidade Federal da 
Para\'{\i}ba\\
Caixa Postal 5008, 58051-970, Jo\~ao Pessoa, Para\'{\i}ba, Brazil}
\email{cffarias,jroberto,petrov@fisica.ufpb.br}

\begin{abstract}
We calculate the one-loop effective potential for 
Horava-Lifshitz-like QED and Yukawa-like theory for arbitrary values of the critical exponent and the space-time dimension.
\end{abstract}
\maketitle
 
\paragraph{Introduction.} The Horava-Lifshitz (HL) methodology based on the essential asymmetry between time and space coordinates \cite{Hor} has been originally introduced within the context of the search for the perturbatively consistent gravity theory. The main advantage of this concept consists in the fact that, from one side, it improves the renormalizability of the field theory models, and, from another side, it avoids arising of the ghosts whose presence is characteristic for the theories with higher time derivatives \cite{Hawk}. Therefore this concept (or, more generally, the concept of the time-space asymmetry) began to be applied not only within studies of gravity but also for the consideration of other (f.e. scalar and vector) field theory models. 

One line of studies of the theories with the time-space asymmetry is devoted to investigation of their renormalizability and renormalization. Within this context, the HL versions of the gauge field theories \cite{ed}, scalar field theories \cite{Anselmi} (see also \cite{Gomes0} for the renormalization group issues), four-fermion theory \cite{ff} and $CP^{N-1}$ model \cite{cpn} were considered. Another important result in this context is the generalization of the Ward identities for the HL-like theories \cite{Gomes}.

Another line of the studies of the HL-like theories is devoted to the effective potential in such theories. In the papers \cite{Eune,Liou,Fara,our} the one-loop effective potential for the scalar field theories with different forms of self-coupling and arbitrary values of $z$, for the scalar QED with $z=2$ and $z=3$, and for the Yukawa model with $z=2$ and $z=3$ has been obtained. However, the interesting problem is the calculation of the (one-loop) effective potential in the scalar QED and Yukawa model with an arbitrary value of the critical exponent. This problem is considered in the paper.

\paragraph{Scalar QED.} The Lagrangian of the scalar QED  with an arbitrary $z$ is
\bea
\label{lasqed}
L&=&\frac{1}{2}F_{0i}F_{0i}+(-1)^z\frac{1}{4}F_{ij}\Delta^{z-1} F_{ij}-D_0\phi
(D_0\phi)^*+
D_{i_1}D_{i_2}\ldots D_{i_z}\phi(D_{i_1}D_{i_2}\ldots D_{i_z}\phi)^*.
\eea
where $D_0=\pa_0-ieA_0$, $D_i=\pa_i-ieA_i$ is a gauge covariant
derivative. For the sake of the simplicity, we suggest that there is no self-coupling of the matter field, the theory is massless, and the critical exponents for scalar and gauge fields are the same (the generalization for the case of their difference is straightforward, as well as for the case of the massive theory).

The propagator for the scalar field has the simplest form
\bea
<\phi\phi^*>=\frac{i}{k^2_0-\vec{k}^{2z}}.
\eea
To find the propagator of the gauge field, we must introduce the gauge-fixing term.
The general form of the gauge-fixing term implies in arising of the complicated matrix propagator mixing the $A_0$ and $A_i$ fields. To simplify the consideration, we must choose the gauge-fixing term allowing for the separation of the quadratic Lagrangians of these fields. 

In the exact analogy with \cite{Fara,our}, the appropriate gauge-fixing term is nonlocal:
\bea
L_{gf}=\frac{1}{2}(-1)^z\Big[(-1)^z\Delta^{-(z-1)/2}\pa_0A_0+\Delta^{(z-1)/2}\pa_iA_i\Big]^2.
\eea
This gauge-fixing term at $z=1$ reduces to the usual, Feynman one.
Adding this gauge-fixing term to the Lagrangian (\ref{lasqed}), we obtain the following
complete quadratic Lagrangian of the gauge field:
\bea
L_c=\frac{1}{2}A_0[\Delta-\frac{\pa^2_0}{(-\Delta)^{z-1}}]A_0-\frac{1}{2}
A_i[\pa^2_0+(-\Delta)^z]A_i.
\eea
So, we succeeded to separate the quadratic actions for the $A_0$ and $A_i$ fields.
The corresponding propagators have rather simple forms:
\bea
\label{props}
<A_0A_0>&=&-\frac{i\vec{k}^{2z-2}}{k^2_0-\vec{k}^{2z}};\nonumber\\
<A_iA_j>&=&-\frac{i\delta_{ij}}{k^2_0-\vec{k}^{2z}}.
\eea

Let us briefly discuss the renormalizability of the theory (\ref{lasqed}). It is well known that the field theory model is renormalizable if the mass dimensions of all couplings in it are non-negative. Since the dimensions of the derivatives $\pa_i$ and $\pa_0$ within the HL formalism are 1 and $z$ respectively, one finds that in the $d$-dimensional space, the dimension of $A_0$ is $\frac{d+z}{2}-1$, and of $A_i$ is $\frac{d-z}{2}$. Therefore the dimension of the coupling $e$ is $\frac{z-d}{2}+1$. As a result, we find that the scalar QED with an arbitrary $z$ is (super)renormalizable in a $d$-dimensional space if $z \geq d-2$. The usual $(3+1)$-dimensional QED, with $z=1$ and $d=3$, is a perfect example of the renormalizable case. If we increase the value of $z$, the renormalization properties of a theory improve. The detailed discussion of renormalizability of the HL-like theories can be found in \cite{ed}.

By the common definition \cite{BO,CW}, the effective action $\Gamma[\Phi]$ is defined as a generating functional of one-particle-irreducible Green functions. It is obtained from the expression
\bea
\label{intmov}
e^{i\Gamma[\Phi]}=\int D\varphi e^{iS[\Phi+\varphi]}|_{1PI},
\eea
where $\Phi$ is a set of background fields, and $\varphi$ is a set of the quantum fields, and the subscript $1PI$ reflects the fact that only one-particle-irreducible contributions are taken into account. In the theories involving fields of different natures, in particular, those ones we consider in this paper, the effective potential depends only on the background matter fields, while the gauge fields (and, in other theories, fields of other natures) are purely quantum ones, and the derivatives of the background fields are put to zero \cite{BO}. 

Then, in the one-loop approximation one must take into account only the vertices associated to two quantum fields (the vertices involving more quantum fields are relevant only in higher loop orders). After ``rationalization'' of their form (that is, moving all derivatives to just one of the quantum fields, cf. \cite{our}), the relevant vetices look like
\bea
\label{racvert}
&&-e^2A_0A_0\Phi\Phi^*;\quad\, -ie(\Phi^*\phi-\Phi\phi^*)\pa_0 A_0,\nonumber\\
&&-ie[\Phi\phi^*-\Phi^*\phi](\pa_j\Delta^{z-1} A_j),\quad\,  
-e^2A_j\Delta^{z-1} A_j\Phi\Phi^*.
\eea
Here $\Phi$ and $\Phi^*$ are the background fields, and $\phi$, $\phi^*$ are quantum ones. We note that the methodology of calculation in the case of an arbitrary $z$ does not essentially differ from the cases $z=2$ and $z=3$ (see \cite{our,Fara}).

As it is usual in different scalar-gauge theories, independently of their nature (see f.e. \cite{GR}), there 
are two possible types of contributions to the one-loop effective potential. In the first of them, depicted at Fig.1, all diagrams represent themselves as a cycles of the gauge propagators only.
The total result of all such diagrams is a sum of two contributions to the
effective potential -- the first one,
$U_a$, is given by sum of loops composed of $<A_0A_0>$ propagators, and the
second one, $U_b$ -- by the sum of loop composed of $<A_iA_j>$ propagators:
\bea
\label{gab}
U_a&=&-\sum\limits_{n=1}^{\infty}\frac{1}{n}\int
\frac{d^dkdk_0}{(2\pi)^{d+1}}
\Big(\frac{2e^2\Phi\Phi^*\vec{k}^{2z-2}}{k^2_0-\vec{k}^{2z}}
\Big)^n=\int\frac{d^dkdk_0}{(2\pi)^{d+1}}\ln[1-
\frac{2e^2\Phi\Phi^*\vec{k}^{2z-2}}{k^2_0-\vec{k}^{2z}}];\nonumber\\
U_b&=&-\sum\limits_{n=1}^{\infty}\frac{d}{n}\int
\frac{d^dkdk_0}{(2\pi)^{d+1}}
\Big(\frac{2e^2\Phi\Phi^*\vec{k}^{2z-2}}{k^2_0-\vec{k}^{2z}}
\Big)^n=d\int\frac{d^dkdk_0}{(2\pi)^{d+1}}\ln[1-
\frac{2e^2\Phi\Phi^*\vec{k}^{2z-2}}{k^2_0-\vec{k}^{2z}}].
\eea
To carry out the sum, we have used the identity 
$\sum\limits_{n=1}^{\infty}\frac{a^n}{n}=-\ln(1-a)$.

\begin{figure}[ht]
\centerline{\includegraphics{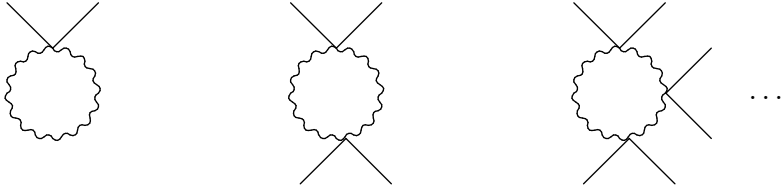}} 
\caption{Contributions involving gauge propagators only.}
\end{figure}

In the second type of diagrams, the matter propagators and triple vertices are present. 
To proceed with the calculations, we should first introduce a "dressed" propagator of the gauge field, see Fig. 2.

\begin{figure}[ht]
\centerline{\includegraphics{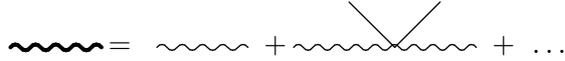}} 
\caption{''Dressed'' propagator of the gauge field.}
\end{figure}

In this propagator, the summation over all quartic vertices is
performed. As a  result, these "dressed" propagators are equal to
\bea
\label{dress}
<A_0A_0>_D&=&<A_0A_0>\sum\limits_{n=0}^{\infty}[2ie^2\Phi\Phi^*<A_0A_0>]^n= 
\frac{i\vec{k}^{2z-2}}{k^2_0-\vec{k}^{2z}-2e^2\vec{k}^{2z-2}\Phi\Phi^*};\nonumber\\
<A_iA_j>_D&=&-\frac{i\delta_{ij}}{k^2_0-\vec{k}^{2z}}\sum\limits_{n=0}^{\infty}
[2e^2\frac{\vec{k}^{2z-2}}{k^2_0-\vec{k}^{2z}}\Phi\Phi^*]^n=
-\frac{i\delta_{ij}}{k^2_0-\vec{k}^{2z}-2e^2\vec{k}^{2z-2}\Phi\Phi^*}.
\eea

The corresponding contribution to the effective potential can be obtained through the summation over all possible one-loop Feynman diagrams representing themselves as cycles of all numbers of links, with each link composed by the product of $<\phi\bar{\phi}>$ and "dressed" $<A_0A_0>_D$ (or $<A_iA_j>_D$) propagators. 
Such diagrams are depicted at Fig. 3.

\begin{figure}[ht]
\centerline{\includegraphics{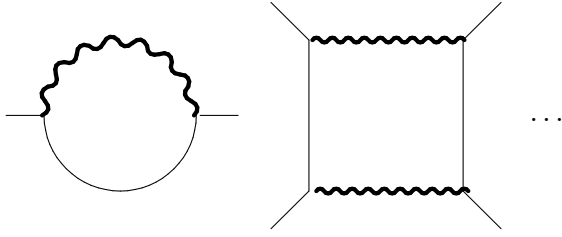}} 
\caption{Contributions involving both gauge and matter propagators.}
\end{figure}

Using the "rationalized" form of the triple vertices
(\ref{racvert}) implies in the need to consider the objects
\bea
G_1&=&<\pa_0A_0(t_1,\vec{x}_1)\pa_0A_0(t_2,\vec{x}_2)>_D;\nonumber\\
G_2&=&<\pa_i\Delta^{z-1} A_i(t_1,\vec{x}_1)\pa_j \Delta^{z-1} A_j(t_2,\vec{x}_2)>_D,
\eea
whose Fourier transforms are
\bea
\label{efpr}
G_1(k)&=&\frac{ik^2_0\vec{k}^{2z-2}}{k^2_0-\vec{k}^{2z}-2e^2\vec{k}^{2z-2}\Phi\Phi^*};
\nonumber\\
G_2(k)&=&-\frac{i\vec{k}^{4z-2}}{k^2_0-\vec{k}^{2z}-2e^2\vec{k}^{2z-2}\Phi\Phi^*}.
\eea
Then, since the effective propagators $G_1$
and $G_2$ enter the diagrams above on the same base, the total
contribution must be symmetric under replacement $G_1\leftrightarrow
G_2$. Hence, the total  contribution from these graphs is
\bea
\label{sumgc1}
U_c&=&-\sum\limits_{n=1}^{\infty}\frac{1}{n}\int
\frac{d^dkdk_0}{(2\pi)^{d+1}}
\Big(2e^2\Phi\Phi^*(G_1+G_2)<\phi\phi^*>\Big)^n,
\eea
which yields
\bea
\label{gc}
U_c&=&-\sum\limits_{n=1}^{\infty}\frac{1}{n}\int
\frac{d^dkdk_0}{(2\pi)^{d+1}}
\Big(-\frac{2e^2\Phi\Phi^*\vec{k}^{2z-2}}{k^2_0-\vec{k}^{2z}-2e^2\vec{k}^{2z-2}\Phi\Phi^*}\Big)^n.
\eea
Carrying out the sum, we arrive at
\bea
\label{corrs}
U_c&=&-\int\frac{d^dkdk_0}{(2\pi)^{d+1}}\ln[1-\frac{2e^2\Phi\Phi^*
\vec{k}^{2z-2}}{k^2_0-\vec{k}^{2z}}].
\eea
So, the $U_c$ completely cancels the $U_a$, cf. (\ref{gab}). It is worth to mention that the cancellation between some contributions is common for the one-loop calculations of the effective potential, see \cite{our,GR}. So, the  effective potential is completely described by $U_b$.
Carrying out the Wick rotation and adding the irrelevant field-independent constant factor 
$id\int\frac{d^dkdk_0}{(2\pi)^{d+1}}\ln[1+\frac{\vec{k}^4}{k^2_0}]$,
we find that the effective potential looks like
\bea
U^{(1)}&=&id\int\frac{d^dkdk_{0E}}{(2\pi)^{d+1}}\ln[1+
\frac{\vec{k}^{2z}+2e^2\Phi\Phi^*\vec{k}^{2z-2}}{k^2_0}].
\eea
Performing the integral along the same lines as in \cite{our}, after returning to the Minkowski space we arrive at 
\bea
U^{(1)}&=&-\frac{d\pi^{\frac{d-1}{2}}}{{4(2\pi)^d}}(2e^2\Phi\Phi^*)^{\frac{d+z}{2}}\frac{\Gamma\Big(-\frac{d+z}{2}\Big)\Gamma\Big(\frac{d+z-1}{2}\Big)}{\Gamma\Big(\frac{d}{2}\Big)}
\eea
This expression can be verified to reduce to that one obtained in \cite{our} for the $z=2$. 
So, the procedure of the calculations does not essentially differ from that case. 

Let us also discuss the physical consequences of this potential. First of all, it can be rewritten as
\bea
U^{(1)}&=&-c(2e^2\Phi\Phi^*)^{\frac{d+z}{2}}\Gamma\Big(-\frac{d+z}{2}\Big),
\eea
where $c=\frac{d\pi^{\frac{d-1}{2}}}{{4(2\pi)^d}}\frac{\Gamma\Big(\frac{d+z-1}{2}\Big)}{\Gamma\Big(\frac{d}{2}\Big)}$ is a positive constant.
There are two characteristic cases.

(i) We have $d+z=2n$ (i.e. either both $z$ and $d$ are odd, as in the usual $(3+1)$-dimensional QED, or both $z$ and $d$ are even). In this case the effective potential displays divergences: if we carry out dimensional regularization supposing that $d+z=2n-\epsilon$ and take into account that $\Gamma(-n+\frac{\epsilon}{2})=\frac{(-1)^n}{n!}(\frac{2}{\epsilon}+\gamma+\lambda_n+O(\epsilon))$, where $\lambda_n$ is a some positive number, we obtain
\bea
U^{(1)}=\frac{(-1)^{n-1}}{n!}c(2e^2\Phi\Phi^*)^n(\frac{2}{\epsilon}+\gamma+\lambda_n-\ln\frac{2e^2\Phi\Phi^*}{\mu^2}).
\eea
This expression, after subtracting of the divergence with a corresponding counterterm, changes the sign depending on $n$. Therefore, for the odd $n$ the effective potential displays the minimum at $\Phi=0$, and for the even $n$ -- at some non-zero $\Phi$. It is easy to see that in the last case the $Z_2$ symmetry $\Phi\to -\Phi$ is broken, thus, the one-loop correction induces the spontaneous symmetry breaking as in \cite{CW}.

(ii) We have $d+z=2n+1$. In this case we have
\bea
U^{(1)}&=&-c(2e^2\Phi\Phi^*)^{n+\frac{1}{2}}\Gamma\Big(-\frac{2n+1}{2}\Big),
\eea
and, since $\Gamma\Big(-\frac{2n+1}{2}\Big)=\frac{(-1)^{n+1}\sqrt{\pi}2^{n+1}}{(2n+1)!!}$, one finds that for the even $n$, the effective potential has a minimum at $\Phi=0$, and there is no symmetry breaking, and for the odd $n$, the effective potential is negative and dispays a maximum at $\Phi=0$, and the theory is unstable.

\paragraph{Yukawa theory.} Then, let us formulate the Yukawa
theory. It is natural to consider now the arbitrary $z$ version of the spinor
field theory, so, the $(d+1)$-dimensional Lagrangian for the theory
looks  like
\bea
\label{yukawa}
L=\bar{\psi}(i\gamma^0\pa_0+(i\gamma^i\partial_i)^z+h\Phi)\psi.
\eea
To study the one-loop effective potential, it is sufficient to treat the scalar field
as a purely external one, and to put the spinor field to be massless since its nontrivial mass implies only in the redefinition of the $\Phi$ field. In this case the loop expansion is closed with the one-loop contribution, and no restriction of the renormalizability of the theory related to the dimension of the coupling $h$ emerges. However, if we suggest that the $\Phi$ is also dynamical, with the same critical exponent $z$ as the $\psi$, that is, is free Lagrangian is the same as in the theory (\ref{lasqed}), the mass dimension of $h$ turns out to be equal to $\frac{3z-d}{2}$. Thus, if one has $z\geq \frac{d}{3}$, the theory is renormalizable -- in particular, it is just the case of the usual, $z=1$ Yukawa model in $(3+1)$-dimensional space. 

The one-loop effective potential corresponding to the Lagrangian (\ref{yukawa}), looks like
\bea
U^{(1)}=-i{\rm Tr}\ln(i\gamma^0\pa_0+(i\gamma^i\partial_i)^z+h\Phi).
\eea
We have two possibilities. In the first of them, the $z$ is even, so, $(i\gamma^i\partial_i)^z=(-\Delta)^{z/2}$, and, following the lines of \cite{our},
we find that the following effective potential is
\bea
U^{(1)}=-i\frac{\delta}{2}\int\frac{d^dkdk_0}{(2\pi)^{d+1}}
\ln\Big[\frac{k^2_0-((\vec{k}^2)^{z/2}+h\Phi)^2}{k^2_0}
\Big].
\eea
Doing the Wick rotation and integrating over $k_0$, we arrive at
\bea
U^{(1)}=-\frac{\delta}{2}\int\frac{d^dk}{(2\pi)^d}((\vec{k}^2)^{z/2}+h\Phi).
\eea
This integral, for any positive $d=3-\epsilon$, vanishes within the dimensional  
regularization being proportional to $\frac{1}{\Gamma(-1-\epsilon)}$
which is zero as $\epsilon\to 0$. Thus, the effective potential in the Yukawa-like theory vanishes for any even $z$.

The second possibility is the case of the odd $z$, $z=2l+1$ so, $(i\gamma^i\partial_i)^z=(-\Delta)^l i\gamma^i\partial_i$. In this case we have
\bea
U^{(1)}=
-i{\rm Tr}\ln(i\gamma^0\pa_0+i(-\Delta)^l  \gamma^i\partial_i+h\Phi).
\eea
If $h\Phi=M$, we can write
\bea
\frac{dU^{(1)}}{dM}=-i{\rm Tr}\frac{1}{i\gamma^0\pa_0+i(-\Delta)^l \gamma^i\partial_i+M}=-i\int\frac{dk_0d^dk}{(2\pi)^{d+1}}\frac{1}{\gamma^0k_0+(k^2)^l  \gamma^i k_i+M}
\eea
Calculating the trace, after the Wick rotation we have
\bea
\frac{d\Gamma^{(1)}}{dM}=M\delta\int\frac{dk_0d^dk}{(2\pi)^{d+1}}\frac{1}{k^2_0+(k^2)^{2l+1}+M^2}=\frac{1}{2}M\delta \int\frac{d^dk}{(2\pi)^d}\frac{1}{[(k^2)^{2l+1}+M^2]^{1/2}}.
\eea
This expression can be evaluated in the same way as in the scalar theory. 
First, we integrate over momenta:
\bea
\frac{dU^{(1)}}{dM}=-\delta\frac{\pi^{d/2-1/2}}{(2\pi)^d(2l+1)}\frac{1}{\Gamma(d/2)}\Gamma\Big(\frac{d}{4l+2}\Big)
\Gamma\Big(-\frac{d}{4l+2}+\frac{1}{2}\Big)M^{\frac{d}{2l+1}},
\eea
Finally, we integrate over $M$ and, restoring the expression for the background field, arrive at the following expression for the effective potential:
\bea
U^{(1)}=-\delta\frac{\pi^{d/2-1/2}}{(2\pi)^d(d+2l+1)}\frac{1}{\Gamma(d/2)}\Gamma\Big(\frac{d}{4l+2}\Big)
\Gamma\Big(\frac{1}{2}\big[1-\frac{d}{2l+1}\big]\Big)(h\Phi)^{\frac{d}{2l+1}+1}.
\eea
We see that this effective potential is finite while $\frac{d}{2l+1}\neq 2k+1$, with $k$ is a some integer number. So, just as in the previous case, we have two situations.

(i) One has $\frac{d}{2l+1}\neq 2k+1$. In this case, the effective potential is finite, and its sign depends on the sign of the $\Gamma\Big(\frac{1}{2}\big[1-\frac{d}{2l+1}\big]\Big)$, which can be positive or negative for different values of $d$ and $l$. If the effective potential is positive, there is no spontaneous symmetry breaking, even if we introduce the classical self-coupling of the scalar field of the form $\lambda\Phi^{2n}$, with $\lambda>0$. However, if the effective potential is negative, the theory is unstable if we have no classical self-coupling of the scalar field, and we can have spontaneous breaking of the $Z_2$ symmetry if such a self-coupling is present since in this case the one-loop corrected effective potential evidently possesses nontrivial minima at $\Phi\neq 0$.

(ii) One has $\frac{d}{2l+1}=2k+1+2\epsilon$, at $\epsilon\to 0_+$. In this case the gamma function displays the poles which change the sign dependently on the parity of the $k$ number, that is, $\Gamma(-k+\epsilon)$, at $\epsilon>0$, is positive for the even $k$ and negative at the odd $k$, and the effective potential is equal to $-a\Gamma(-k-\epsilon)\Phi^{2k+2+2\epsilon}$, with $a$ is a positive number. Expanding this expression in power series around $\epsilon=0$, we find that the effective potential is proportional to $(-1)^k$, hence it has minima at $\Phi=0$ for the even $k$, and minima at some other $\Phi=\Phi_0$ for the odd $k$. In other words, the situation is rather similar to the case of the scalar QED, taking into account the fact the different statistics of the quantum fields.

\paragraph{Conclusion.} We studied the one-loop effective potential for two HL-like theories with the arbitrary values of the critical exponent: first, the scalar QED, and second,  the Yukawa theory, where
the one-loop effective potential was shown to vanish for the even $z$, being nontrivial for the odd values of $z$. In certain cases, the effective potential is one-loop finite in both theories. In principle, we
can also introduce the coupling between the gauge and spinor
fields, which, however,  will start to contribute to the effective
potential only at the two-loop order.
One can observe that the
methodology for calculating the effective potential does not
essentially differ from that one in the usual, Lorentz invariant field theories. We emphasize that we succeeded to find the one-loop effective potential for the theories with an arbitrary critical exponent despite the fact that the classical action of the HL-like QED with an arbitrary critical exponent is in general extremely complicated and involves an unlimited number of terms. Also, we showed that in certain cases (that is, the models with a nontrivial the self-coupling of the scalar field, or the situations when the effective potential requires a renormalization) the effective potential can generate a spontaneous $Z_2$ symmetry breaking. We note that, for the case of the QED, these results can be straightforwardly generalized for the $U(N)$ multiplets of arbitrary number scalar fields, so, we can observe a dynamical $U(N)$ symmetry breaking in Horava-Lifshitz-like theories.

{\bf Acknowledgements.} This work was partially supported by Conselho
Nacional de Desenvolvimento Cient\'{\i}fico e Tecnol\'{o}gico (CNPq).
A. Yu. P. has been supported by the CNPq project 303461-2009/8.

\end{document}